\documentstyle[prl,aps,epsfig]{revtex}
\begin{document}
\topmargin  = -1mm
\hoffset = 2mm
\voffset = -5mm
\def\be{\begin{equation}}
\def\ee{\end{equation}}
\def\d{\partial}
\def\bra{\langle}
\def\ket{\rangle}
\title{Asymmetric Cellular Automaton Modeling Earthquakes}
\author{P.G.Akishin$^{1,2,3}$, M.V.Altaisky$^{1,3,4}$, I.Antoniou$^{2,5}$, 
A.D.Budnik$^{1,2}$, V.V.Ivanov$^{1,2}$\\ }
\address{$^1$ Laboratory of Computing Techniques and Automation, 
         Joint Institute for Nuclear Research, Dubna, 141980, RUSSIA \\
$^2$ International Solvay Institutes for Physics and Chemistry, 
         CP-231, ULB, Campus Plaine, Bd. du Triomphe, 1050,\\
          Brussels, BELGIUM \\
$^3$ European Commision, Joint Research Centre, I-21020 Ispra (Va), Italy\\
$^4$ Space Research Institute RAS, Profsoyuznaya 84/32, 
         Moscow, 117810, RUSSIA \\
$^5$ Theoretische Natuurkunde, Free University of  Brussels,  
         Brussels, BELGIUM}
\maketitle

\begin{abstract}
We propose an asymmetric modification of the sand-pile-like 
cellular automaton for earthquake modeling. The cumulative 
event distribution is shown to be dependent on 
the asymmetry parameter. 
\end{abstract}

\section{}

The ideas of self-organized criticality (SOC) and the observation 
of SOC-like behavior of simple dynamical systems, like 
cellular automata (CA), have significantly warmed the 
interest in the application of a CA to earthquake simulation. 
A number of papers on the construction of CA counterpart of 
the well known Burridge-Knopoff (BK) model earthquake faults \cite{BK67} 
have been published \cite{BakTang89,Nak91}. 
Both, the P.Bak \& C.Tang model  \cite{BakTang89} of the complete 
stress release and the H.Nakanishi model \cite{Nak91} with more 
elaborated stress relaxation function show the power law behavior 
which closely resembles the Gutenberg-Richter power law. Both of them 
are evidently too simplistic in comparison to the BK system of 
differential equations for the spring-block model.
What can be done to make the existing CA models closer to the 
BK model? In this paper we present a modification of the existing CA 
which partially answers this question. 

\section{}

One of the basic principles of the construction of cellular automata 
for physical applications consists in conserving the symmetry of the 
original physical system as precisely as possible. Let us compare the 
CA models \cite{BakTang89,Nak91} and the BK spring-block model in this 
aspect. 
\begin{itemize}
\item The BK system is a system of $N$ blocks of mass 
$m_i, i=\overline{1,N}$ resting on a rough surface and connected 
to each other by harmonic springs of stiffness $k_c$; each 
block is attached by a leaf spring of stiffness $k_p$ to the 
moving upper line, see Fig.~1. 

\vskip5mm
\begin{figure}[h]
{\centering \epsfig{file=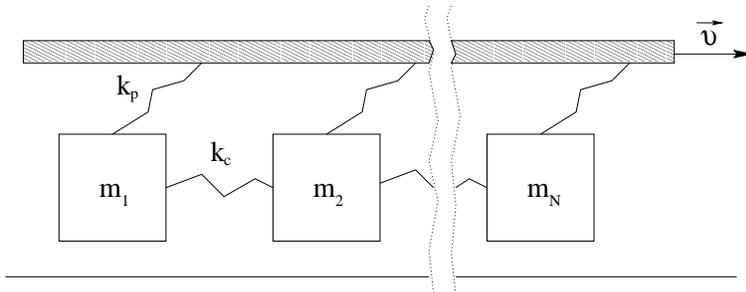, width=10cm} }
\label{bk:pic}
\caption{\small The geometry of the BK model: the system is composed
of $N$ identical blocks of mass $m_i$, $k_c$ is the  stiffness of 
the ``horizontal'' springs,  $k_p$ is the stiffness of the pulling 
springs,  $v$ is the  constant velocity of the pulling line.}
\end{figure}

Initially (at $t=0$) the system is at rest, and the elastic energy 
accumulated in ``horizontal'' springs is only due to randomly generated 
small initial displacements of the blocks from their neutral positions. 
The moving upper line, which simulates the movement of an external driving 
plate, exerts a force $f_n=-k_p(x_n-vt)$ on each $n$-th block. The 
nonlinear friction is defined in such a way that it holds 
each block at rest until the sum of all the forces  applied to 
this block exceeds a certain critical value $F_0$. Then the block makes 
a slip inhibited by nonlinear friction to a new position. A pause 
between two slips is believed to account for a pause between
earthquakes. 
\item The CA of P.Bak and C.Tang for artificial earthquake simulation 
 \cite{BakTang89} is a simple sandpile-like system which obeys 
the following rules:
\begin{enumerate}
  \item The array of $N$ values (``blocks'') is initiated by 
    certain randomly generated values $f^0_i,i=\overline{1,N}$
    (the upper indices are used for discrete time).
  \item If the sum of the forces on $i$-th block exceeds a certain threshold 
        value (usually $f_{Th}=1$, without loss of generality), 
        then ``the accumulated stress'' is shared with 
        the nearest neighbors according to the rule
        \be
        f_{i\pm1} \to f_{i\pm1} + \frac{k_c}{2k_c+k_p}\delta f_i,
        \label{esh}
        \ee
        where $\delta f_i = f_i-f_i'$ is the stress drop of the 
        over-threshold $i$-th block, which is evaluated according 
        to the law
        \be f_i' = \phi(f_i-f_{Th}). \label{relax} \ee
        The relaxation function $\phi$ is model dependent: 
        it may be taken to be zero as in the P.Bak \& C.Tang model, 
        or it may be some decreasing function as in 
        the Nakanishi cellular automaton. 
  \item When the evolution is completed the forces applied to all the 
        blocks are incremented by the tectonic force 
        \be
        f_i' \to f_i' + k_p v \Delta t,\quad  
        i=\overline{1,N}\label{pump}
        \ee   
\end{enumerate}  
\end{itemize}

Now let us turn to the symmetries of the BK and CA models.
Due to the presence of the tectonic driving force $k_p\vec v t$, which 
implies the existence of a preferable space direction, namely, the direction 
of the moving plate velocity $\vec v$, the BK system is apparently not 
invariant under inversion $\vec x \to -\vec x$. The stress 
redistribution law (\ref{esh}) of the CA, in contrast, is completely 
symmetric with respect to this inversion ($f_{i+1}\leftrightarrow f_{i-1}$). 
Therefore, the above mentioned CA has more symmetries than  
the parent BK model. What can be done about this? 
The answer is evident. An asymmetry should be introduced in the stress 
redistribution law (\ref{esh}) as follows: 
\be
f_{i+1} \to f_{i+1} + (1-\gamma)\alpha\delta f_i,\quad
f_{i-1} \to f_{i-1} +    \gamma \alpha\delta f_i.
\label{assh}
\ee
The value of the asymmetry parameter $\gamma =1/2$ 
corresponds to equal sharing (\ref{esh}), $\gamma = 0$ 
leads to completely asymmetric sharing. The factor 
$\alpha=\frac{2k_c}{2k_c+k_p}$ is chosen to comply with \cite{Nak91,OFC92}.

We have performed computer simulations with one-dimensional 
35 blocks CA for different values of the asymmetry parameter 
$\gamma=0.25,0.4,0.45,0.5$. The cumulative event distribution, 
i.e. the number of events of magnitude not less than a given value,  
which is often expected to have the form of the Gutenberg-Richter law,
is presented in Fig.2. 
\begin{center}
\begin{figure}[ht]
{\centering \epsfig{file=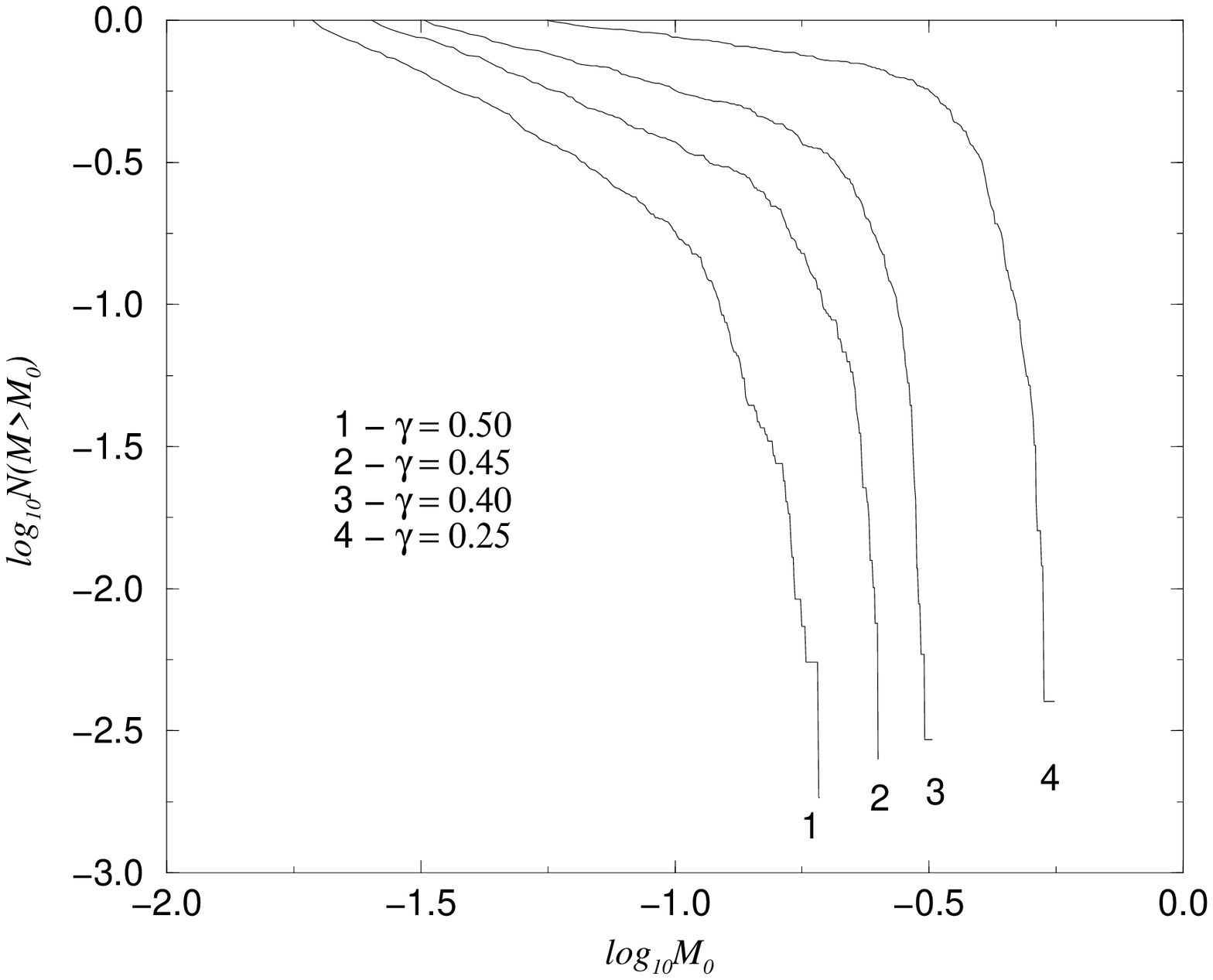, width=10cm} }
\caption{Cumulative event distribution $N(M>M')$ for one-dimensional 35 cell 
         CA calculated for different values of the asymmetry parameter 
         $\gamma=0.25,0.4,0.45,0.5$ ($\gamma=0.5$ corresponds to the
         symmetric case) plotted in logarithmic 
         coordinates vs. the magnitude $m=\lg M$ with 
         the event size understood to be the total stress relaxation 
         $M=\sum_i (f_i-{f_i}')$. The significance level was 
         taken as 0.1 of the maximal magnitude for each run. 
         The relaxation function $\phi$ was taken as in the 
         above cited Nakanishi's paper.}
\label{asym2d:pic}
\end{figure}  
\end{center}  
As it can be seen from the picture, the asymmetry parameter $\gamma$, 
significantly affects the slope of the curve (the logarithm of 
the cumulative event number vs. the magnitude). 
We also observed the amplitude of events to increase with  
asymmetry increasing. This means that $\gamma$ is 
a control parameter of the system, the appropriate choice of which 
can tune the system close or far from the realistic value of 
$b\approx1$ in the Gutenberg-Richter law\cite{GR44} 
$$\lg N(M>M')=a-mb,\quad m=\lg M = \lg\sum_i (f_i-{f_i}').$$

\section*{Acknowledgments}
The authors are grateful to Drs G.~Molchan and M.Shnirman 
for useful discussions and to Prof. I.~Prigogine for his 
enthusiastic support.  

This work was supported by the European Commission through the 
collaboration of the International Solvay Institutes 
with the Joint Research Center, Ispra. 
\bibliographystyle{alpha}
\bibliography{earth}
\end{document}